\documentclass[prb,aps,showpacs,twocolumn]{revtex4}
\usepackage{epsf}
\usepackage[dvips]{color}
\usepackage{graphicx}

\begin{document}

\title{Millimeter Wave Localization: Slow Light and Enhanced Absorption}
\author{John A. Scales, L. D. Carr and D. B. McIntosh}
\affiliation{Department of Physics, Colorado School of Mines, Golden, CO 80401 USA}
\author{V. Freilikher$^1$ and Yu.\,P. Bliokh$^2$}
\affiliation{$^1$ Department of Physics, Bar-Ilan University, Ramat-Gan 52900, Israel\\
$^2$ Physics Department, Technion-Israel Institute of Technology, Haifa
32000 Israel}
\date{\today}

\begin{abstract}
  We exploit millimeter wave technology to measure the reflection and
  transmission response of random dielectric media.  Our samples are
  easily constructed from random stacks of identical, sub-wavelength
  quartz and Teflon wafers. The measurement allows us to observe the
  characteristic transmission resonances associated with localization.
  We show that these resonances give rise to enhanced attenuation even
  though the attenuation of homogeneous quartz and Teflon is quite
  low.  We provide experimental evidence of disorder-induced slow
  light and superluminal group velocities, which, in contrast to
  photonic crystals, are not associated with any periodicity in the
  system.  Furthermore, we observe localization even though the sample
  is only about four times the localization length, interpreting our
  data in terms of an effective cavity model.  An algorithm for the
  retrieval of the internal parameters of random samples (localization
  length and average absorption rate) from the external measurements
  of the reflection and transmission coefficients is presented and
  applied to a particular random sample. The retrieved value of the
  absorption is in agreement with the directly measured value within
  the accuracy of the experiment.
\end{abstract}

\maketitle

\section{Introduction}
\label{sec:introduction}

Multiple scattering of electrons and photons in disordered systems
gives rise to many important physical effects, including Anderson
localization, quantum corrections to conductivity,
localization-induced metal-insulator transitions, and coherent
backscattering~\cite{lifshits,sheng,akkermans,gantmakher}. 
Effects of disorder are most strongly
exhibited in one-dimensional (1D) systems. The combination of
localized resonances and bandgaps in randomly layered samples opens
up fresh opportunities for applications
%(as an alternative to photonic crystals)
and allows for detailed studies of slow light and other anomalous
group velocity effects as well as the enhanced attenuation
associated with the long path length of localized photons. Due to
the ease of fabrication, there are many potential applications for
ordered and disordered millimeter wave dielectrics, including
fundamental physics studies which can exploit the long dwell times
associated with localized photons.

\begin{figure}[t]
\begin{center}
\resizebox{70mm}{!}{\includegraphics{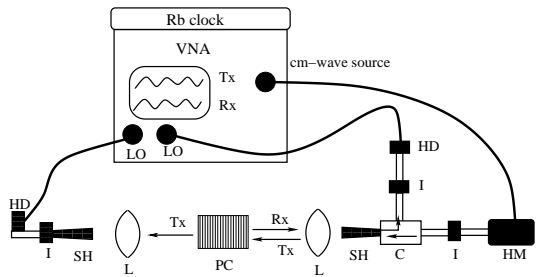}}
\end{center}
\caption{\textit{Experimental setup.} The Vector Network Analyzer
(VNA) generates centimeter waves which pass through a harmonic
multiplier (HM), converting them to millimeter waves. Other
components include isolators (I), a circulator (C), polyethylene
lenses (L), harmonic detectors (HD), scalar (corrugated) horns (SH)
and two phase-locked local oscillators (LO). The thick lines refer
to coaxial cables, which are only used for the cm waves. The
internal frequency synthesizer is locked to a Rubidium clock,
allowing for long-term stability. Finally, the vector receiver in
the VNA down-converts the time-varying E-fields (Tx for transmitted
and Rx for reflected) to a low frequency for digitization and
display. The sample (PC) is at the focal plane of the transmitting
horn.  The thick black lines connecting the VNA and the millimeter wave
components represent high-frequency micrwoave coaxial cables.  
Between the harmonic mixer/detectors and the 
scalar horn antennae, the millimeter waves propagate via waveguide.} 
\label{fig_setup}
\end{figure}

In this paper, we describe observations of Anderson localization,
enhanced absorption, and slow light in millimeter waves 
propagating in a random
dielectric. Specifically, we utilize 100-layer dielectric stacks of
quartz and Teflon wafers randomly shuffled like a deck of cards. The
behavior of wave propagation in this system is a mixture of band
gaps in which the random dielectric becomes an almost perfect
reflector, resonances which are associated with anomalously high
transmission, localized electric field
inside the layered system, and
enhanced absorption.
The block diagram of our experiment is
shown in Fig.~\ref{fig_setup}. A key aspect of our work is that we
achieve these disorder-induced defects in a finite system which is
only about four times the localization length. 
Such finite size
considerations are vital in quantum studies of disorder-induced
phenomena in quantum degenerate ultracold
atoms~\cite{clement2005,fort2005,schulte2005,paul2007}.

There are several approaches to experimental realization of 1D
systems. Recently, dielectric multilayer \cite{lasers, bertolotti}
 and single mode structures \cite{Shapira,kuhl:633, bliokh97}
have been used to study localization in optical and microwave
systems. However, millimeter waves offer several advantages. First,
we can measure both amplitude and phase of the reflected and
transmitted fields simultaneously. Second, our setup is free-space
and quasi-optical, so the attenuation in the waveguide is minimized.
Unlike previous work on photonic crystals and optical
resonators where superluminal and slow light have also been observed 
\cite{photocrys, anomdisp, spielmann, baldit, heebner, yariv}, we
induce these effects via disorder. Ideally, a system would have
sufficiently low attenuation that both the reflection and
transmission response of a disordered system could be measured;
would allow for amplitude and phase of the reflected and transmitted
signal to be measured simultaneously; have large enough bandwidth that a
number of band-gaps and resonances could be studied for each sample;
and would allow for easy fabrication of samples, so that many could
be investigated quickly. All these requirements are met by our
quasi-optical millimeter wave system.

Our article is outlined as follows.  In Sec.~\ref{sec:experiment}, the experimental 
set-up is described.
In Sec.~\ref{sec:data} the transmission and reflection data are presented.
In Sec.~\ref{sec:theory}, the effective cavity model is derived.  In Sec.~\ref{sec:discussion},
the experimental data of Sec.~\ref{sec:data} are compared both to the effective 
cavity model and to simulations
obtained from a propagator matrix.
Finally, in Sec.~\ref{sec:conclusions}, we conclude.

\section{Description of The Experiment}
\label{sec:experiment}

Our layered dielectric samples consist of stacks of 300-400 $\mu$m
thick wafers of fused quartz and Teflon held together tightly in a
telescope tube with retaining rings. These disks are stock items for
Teflon and semiconductor manufacturers and are available at relatively low
cost. With our millimeter wave vector network analyzer, a scan over
the entire W-band (75-110 GHz) at 5000 frequency points takes only a
few minutes. Our setup allows for high signal-to-noise measurement
of both reflected and transmitted E-fields simultaneously. The
electric field is a linearly polarized Gaussian beam whose diameter
is much smaller than that of the wafers; the beam strikes the
dielectric stack perpendicularly, allowing a 1D treatment.

Quartz and Teflon are low loss materials at millimeter wave
frequencies and provide a reasonable dielectric contrast. The real
part of the dielectric constants are measured to be $\epsilon
^{\prime }=3.80 \pm .02 $ for quartz and $\epsilon ^{\prime }=2.05 \pm .02$ for
Teflon; these values, obtained by fitting the transmission data for
homogeneous samples with a Fabry-Perot
model, are quite close to reference values (3.8 and 2.1, respectively~\cite%
{goldsmith}). The same fitting procedure gives loss tangents of
$0.0007 \pm 0.0002$ for quartz and $0.0003 \pm 0.0002$
for Teflon. Such small loss tangents are
difficult to measure with our setup and are less accurately 
specified.

The
micrometer-measured thicknesses of our individual wafers are
$0.460\pm 0.01$ mm for quartz and $0.375\pm 0.005$ mm for Teflon.
However, even though the wafers are held  tightly in the telescope
tube, inevitably there are small air gaps between the wafers. This
ultimately limits our ability to fit the data.

The measurements were performed with a vector network analyzer
developed by AB Millimetre. The millimeter waves are generated by a
sweepable centimeter wave source, i.e., microwaves, in this case
from 8-18 GHz.  For work in the W-band these centimeter waves are
harmonically multiplied by Schottky diodes, coupled into a
waveguide, and eventually radiated into free space by a scalar 
horn.  These are also known as corrugated horns and have the property
that when the TE01 rectangular waveguide mode is coupled into the
circular horn, a hybrid mode is established which has no cross-polarization.
This results in the E-field being vertically polarized in a single-mode
Gaussian beam.\cite{goldsmith}.

A polyethylene lens collimates the beam. The random dielectric stack is
placed in the focal plane of the quasi-optical system. The
transmitted field is then collected by an identical lens/horn
combination, detected by a Schottky harmonic detector and fed to a
vector receiver which mixes the centimeter waves down to more easily
manageable frequencies where the signal is digitized. Reflected
waves are also collected by the transmitting horn and routed via a
circulator and isolator to the vector receiver. The source and
receiver local oscillators are phase-locked.  A more
complete description of the system is given in \cite{mvna:rsi} and 
\cite{mmwspectroscopy}.

\section{Experimental Data}
\label{sec:data}

\begin{figure*}
\centerline{
\includegraphics[width=150mm]{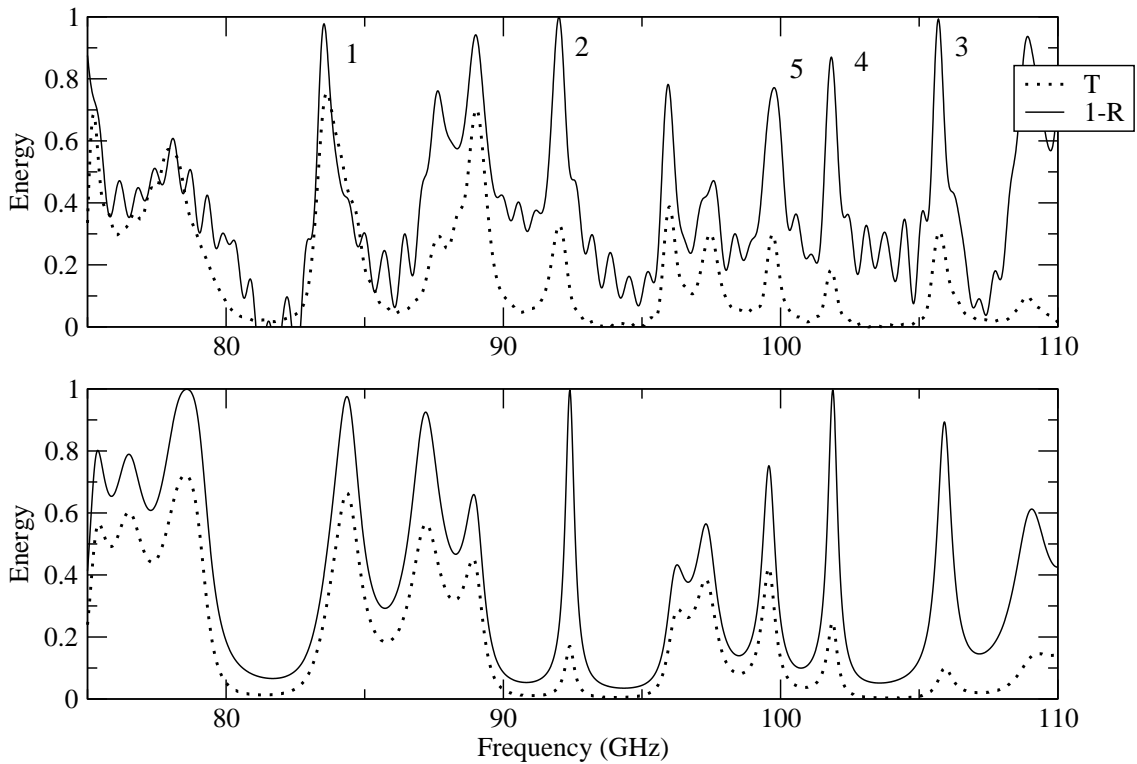}
}
\caption{\textit{Measured (top) and calculated (bottom) transmission
    and reflection response for a randomly layered sequence of 100
    quartz and Teflon disks.} Plotted are the transmission coefficients ($T$,
    dotted curves) and one minus the reflection coefficient ($1-R$, solid curves).
  }
\label{fig_trans}
\end{figure*}

In Fig.~\ref{fig_trans}, we illustrate the transmission and
reflection spectra for the W-band as investigated experimentally and
theoretically for a representative sample. 
\footnote{ This particular sample has the following pseudo-random sequence
of quartz (Q) and Teflon (T) wafers:
Q T Q T T Q Q T T Q T Q Q T T Q Q Q T T Q T T Q Q Q Q Q T T T Q Q T T T T Q Q 
T Q Q T Q Q Q T T Q T Q Q Q T Q Q T T Q Q T Q Q T T Q T Q T 
Q T T Q Q T Q Q Q Q T Q T T T T Q T Q Q Q Q Q T T T Q T T T T.}
Plotted are the transmission coefficient ($T$) and one minus the reflection coefficient
(R).

The actual data collection requires careful calibration of the
system, in the instrument response and in all phase shifts in the
cables and waveguide. First, one sweeps over the band of frequencies
with no sample present; this gives a calibration E-field 
$E_{0}^{\mathrm{t}}$ measured for transmission. Second, one sweeps
with a ``perfect'' reflector, i.e., a polished metal plate, at the
focal plane; this gives a calibration
E-field $E_{0}^{\mathrm{r}}$. Finally, with a sample in
place, the sweep returns the amplitude and phase due to the presence
of the sample relative to a flat instrument response. So, in
Fig.~\ref{fig_trans}, $R=|E^{r}/E_{0}^{r}|^{2}$ and
$T=|E^{t}/E_{0}^{t}|^{2}$ are the reflection and transmission
coefficients, respectively.

The theoretical curve in Fig.~\ref{fig_trans} is obtained by solving
Maxwell's equations numerically using the standard propagator matrix
approach~\cite{li:046607}. This allows us to compute the amplitude
and phase of the reflected and transmitted fields as well as the
fields everywhere inside the sample. 
% The transmitted phase $\phi
% _{t}$ is given by the sum of the unwrapped phase of the transmission
% amplitude and an integer multiple of $\pi $ determined
% uniquely such that the phase will vanish as $\omega \rightarrow 0$~\cite%
% {photocrys}. 
As can be seen from Fig.~\ref{fig_trans}, most
frequencies lie in disorder-induced band-gaps, where the reflection
coefficient is near unity. Between band-gaps there are narrow
resonances accompanied by greatly enhanced transmission. 
Off-resonance transmission losses, e.g., around 80 and 94 GHz in Figure
\ref{fig_trans},
are about 1\%.

We observe in Fig.~\ref{fig_trans} differences in both amplitude and
frequency between the measured and calculated spectra.  In order to
understand this effect, we studied periodic systems (data not
shown), and found that the fit  was better.  We also studied
many different random samples and did not find any systematic
amplitude and frequency shifts.  We conclude that these shifts are 
likely a
random effect of the tiny, unmeasurable air gaps between wafers as well
as of small random variations in the thicknesses of the wafers.

From the data  shown in Fig. \ref{fig_trans}, the localization length $l_{
\mathrm{loc}}$ can be estimated. Indeed, it is well known that
the Lyapunov exponent, $\gamma \equiv l_{\mathrm{loc}}^{-1}=-(\ln T)/L$
is a self-averaging quantity~\cite{lifshits,sheng,akkermans,gantmakher}. 
That is, its fluctuations are
small and its variance tends to zero, so that 
\begin{equation}
\gamma \rightarrow \langle \gamma
\rangle
\ge -(1/l_{\mathrm{loc}}+1/l_{\mathrm{abs}})\,,
\end{equation}
where $L$ is
the sample length and $l_{\mathrm{abs}}$ is the spatial absorption
rate \cite{PhysRevB.50.6017}. Thus, at large $L$  the Lyapunov exponent 
$\gamma$ is close to $-(1/l_{\mathrm{loc}}+1/l_{\mathrm{abs}})$
with high probability, i.e., in almost all samples, or at almost
all frequencies in one sample. We call such off-resonance samples and frequencies
\emph{typical}. 

In Fig.~\ref{fig_trans}, 
we find that $\gamma L$ at typical
frequencies is approximately  $0.01$.  The measured loss
tangents of quartz and Teflon correspond to absorption lengths of
approximately 35 cm for quartz and 100 cm for Teflon.  
The reciprocal absorption length is $2\pi n\tan \delta /\lambda $, where 
$\tan \delta =\epsilon ^{\prime \prime }/\epsilon ^{\prime }$, $n$ is the
index of the homogeneous material and $\lambda $ is the free-space
wavelength of the probing beam.  
The absorption lengths of both quartz and Teflon
are much larger than the
sample itself.  Thus, we find for our 42 mm sample the localization length $l_{
\mathrm{loc}}=\gamma ^{-1}\simeq 10$  mm. Since  $l_{\mathrm{abs}}>>L$,
its influence on typical transmission frequencies
is negligible, 
\begin{equation}T_{\mathrm{typ}}\approx \exp (-L/l_{\mathrm{loc}}).
\label{eqn:localizationlength}
\end{equation}
We reserve discussion of the atypical, or \emph{resonant} case 
to Secs.~\ref{sec:theory} and~\ref{sec:discussion}.

\section{Theoretical Description}
\label{sec:theory}

As can be seen from Fig.~\ref{fig_trans}, the sum of the reflection
and transmission coefficients drops noticeably at the resonances.
This means that absorption plays a 
significant role; in the presence of absorption, standard analytical
treatments~\cite{sheng,akkermans,gantmakher} fall short.  Therefore,
we take advantage of an effective cavity model recently introduced by
Bliokh et al.~\cite{bliokh97,bliokh2004} as follows.  In the random
medium of our sample, the transmission coefficient at typical
frequencies determines the localization length via
Eq.~(\ref{eqn:localizationlength}) and so can be measured.
However, along with the typical frequencies there are 
\emph{resonant} frequencies.
These are relatively rare, or low
probability frequencies at which resonances occur, and the effect of
absorption is dramatic.  
In these resonances, photons become partially
trapped by multiple scattering.
%, resulting in very low group velocity.
In effect, the random multiple scattering gives rise to an effective
cavity within the sample.
The long dwell time of these photons
is responsible for the enhanced attenuation. In other words, the
attenuation increases significantly at a resonance because the
energy absorbed at each point is proportional to the intensity of
the field, which is exponentially large throughout the resonant
cavity. Therefore, both the reflection and transmission coefficients
decrease sharply even when the absorption length $l_{\mathrm{abs}}$
is much larger than the total length $L$ of the sample.

One can treat any particular localized state as an effective cavity
mode.  Using the measurable reflection and transmission coefficients
on and off resonance together with the measurable resonance width, one
can make a fit to determine the location and width of the effective
cavity, as well as an effective absorption length.  

Consider a random
sample as a 1D resonator built of a cavity of length
$l_{\mathrm{cav}}$ bounded by two barriers of lengths $l_{1}$ and
$l_{2}$, whose transmission coefficients are
\begin{equation}T_{1,2}=\exp
  (-l_{1,2}/l_{\mathrm{loc}})\,,\label{eqn:t12defn}\end{equation} with
$l_{\mathrm{loc}}$ the localization length.  The parameters of the
effective cavity will be different for each resonance inside the same
sample.  We take as our starting point the resonant transmission and
reflection coefficients and resonant width, which can be shown to be
\begin{eqnarray}
R_{\mathrm{res}} &=&1-{\frac{4}{\left( \sqrt{u/v}+\sqrt{v/u}\right) ^{2}}}\,,  \label{eqn:start1} \\
T_{\mathrm{res}} &=&{\frac{4}{(u+v)^{2}}}\,,  \label{eqn:start2} \\
\delta \omega &=&\frac{c}{2\bar{\varepsilon} \,l_{\mathrm{cav}}}\sqrt{T_{1}T_{2}}(u+v)\,,
\label{eqn:start3}
\end{eqnarray}
respectively, where $c$ is the speed of light in vacuum, 
$\bar{\varepsilon}$ is the average real part of the dielectric
constant and $\delta \omega = 2\pi \delta f$ is the resonance width.  
Here
\begin{eqnarray}
u&\equiv&\sqrt{{\frac{T_{1}}{T_{2}}}},\label{eqn:def1}\\
v&\equiv&{\frac{G}{\sqrt{T_{1}T_{2}}}}+\sqrt{{\frac{T_{2}}{T_{1}}}}\,,\label{eqn:def2}\\
G&\equiv&2\sqrt{{\bar{\varepsilon} }}\,l_{\mathrm{cav}}/l_{\mathrm{abs}}\,,\label{eqn:def3}
\end{eqnarray}
where $l_{\mathrm{abs}}$ is the effective 
absorption length such that the intensity 
$I$ decays as $\exp (-x/l_{\mathrm{abs}})$ in an ordinary random sample, 
and the total length of the sample is 
\begin{equation} L=l_{1}+l_{2}+l_{\mathrm{cav}}\,. 
\label{eqn:totallength}\end{equation}

First, we determine the effective cavity length $l_{\mathrm{cav}}$.  
By definition, 
\begin{equation}
T_{1}T_{2}=e^{-(L-l_{\mathrm{cav}})/l_{loc}}.  \label{eqn:int3}
\end{equation}%
Substituting Eq.~(\ref{eqn:int3}) and Eq.~(\ref{eqn:start2}) into Eq.~(\ref{eqn:start3}), 
one finds an implicit equation for the effective cavity length: 
\begin{equation}
l_{\mathrm{cav}}=\frac{c}{\bar{\varepsilon} \,\delta \omega } \frac{1}{\sqrt{T_{\mathrm{res}}}}
e^{-(L-l_{\mathrm{cav}})/(2l_{\mathrm{loc}})}\,.  \label{eqn:cavitylength}
\end{equation}
This transcendental equation is analyzed in more detail in Sec.~\ref{sec:discussion}.

Second, we determine the effective absorption length $l_{\mathrm{abs}}$.  
From Eqs.~(\ref{eqn:def1})-(\ref{eqn:def3}),
\begin{equation}
\frac{2\sqrt{\bar{\varepsilon}}\,l_{\mathrm{cav}}}{l_{\mathrm{abs}}}=\sqrt{T_{1}T_{2}}\left( v-u^{-1}\right)\,.
\end{equation}
Then Eq.~(\ref{eqn:int3}) implies
\begin{equation}
l_{\mathrm{abs}}=2\sqrt{\bar{\varepsilon}}\,l_{\mathrm{cav}}\left( v-u^{-1}\right)^{-1}
e^{(L-l_{\mathrm{cav}})/(2l_{\mathrm{loc}})}.  \label{eqn:int4}
\end{equation}%
Let us solve for $u,v$ as functions of $R_{\mathrm{res}},T_{\mathrm{res}}$ alone.
Equations~(\ref{eqn:start1})-(\ref{eqn:start2}) yield
\begin{eqnarray}
u^{2}+2uv+v^{2} &=&\frac{4uv}{1-R_{\mathrm{res}}},  \label{eqn:int1a} \\
u^{2}+2uv+v^{2} &=&\frac{4}{T_{\mathrm{res}}},  \label{eqn:int1b}
\end{eqnarray}
It follows that
\begin{eqnarray}
v=\frac{1 \pm \sqrt{R_{\mathrm{res}}}}{\sqrt{T_{\mathrm{res}}}},  \label{eqn:int2a}\\
u=\frac{1 \pm \sqrt{R_{\mathrm{res}}}}{\sqrt{T_{\mathrm{res}}}},  \label{eqn:int2b}.
\end{eqnarray}
Equations~(\ref{eqn:int2a}) and~(\ref{eqn:int2b})
are valid only for $R_{\mathrm{res}}\neq 1$ and $T_{\mathrm{res}}\neq 0$.
Substitution of Eqs.~(\ref{eqn:int2a}),~(\ref{eqn:int2b}), and~(\ref{eqn:cavitylength}) 
into Eq.~(\ref{eqn:int3}) yields the absorption length:
\begin{equation}
l_\mathrm{abs} \equiv \frac{1}{\Gamma} = 
2 \frac{c}{\delta \omega \sqrt{\bar{\varepsilon}}}
\frac{1 \pm
\sqrt{R_{\mathrm{res}}}}{1-R_{\mathrm{res}}-T_{\mathrm{res}}}.  \label{eqn:absorptionlength}
\end{equation}

Third, we determine the length of the first barrier of the cavity, $l_1$.
Substituting Eqs.~(\ref{eqn:t12defn}) and~(\ref{eqn:totallength}) 
into Eq.~(\ref{eqn:def1}), one obtains 
\begin{equation}
u=e^{-l_{1}/l_{loc}}e^{(L-l_{\mathrm{cav}})/(2l_{\mathrm{loc}})}\,.
\end{equation}
Eliminating $u$ with Eq.~(\ref{eqn:int2b}) and solving for $l_{1}$, 
\begin{equation}
l_{1}={\frac{1}{2}}(L-l_{\mathrm{cav}})-l_{\mathrm{loc}}\ln
\frac{1 \pm \sqrt{R_{\mathrm{res}}}}{\sqrt{T_{\mathrm{res}}}}.
\label{eqn:int5}
\end{equation}%
Finally, substituting Eq.~(\ref{eqn:cavitylength}) into Eq.~(\ref{eqn:int5}), one finds
the length of the first barrier:
\begin{eqnarray}
l_{1}=-l_{\mathrm{loc}}\ln \left[ {\frac{\bar{\varepsilon} \,\delta \omega \,l_{\mathrm{cav}}}{c}}
\left( 1 \pm \sqrt{R_{\mathrm{res}}}\right) \right] .  \label{eqn:firstbarrierlength}
\end{eqnarray}

Equations~(\ref{eqn:absorptionlength}),~(\ref{eqn:cavitylength}),
and~(\ref{eqn:firstbarrierlength}) present the absorption length
and the effective cavity lengths in terms of measurable quantities,
namely, the resonance transmission and reflection coefficients
$T_{\mathrm{res}}$ and $R_{\mathrm{res}}$, the resonance width
$\delta\omega=2\pi \delta f$, the average dielectric constant
$\bar{\varepsilon}$, and the localization length determined via
Eq.~(\ref{eqn:localizationlength}) from off-resonant transmission
coefficients.  In this treatment we have not taken advantage of the
extra phase information that is obtained by our system.  The phase provides
an extra datum at each frequency which can be used to obtain additional
cavity parameters.

Finally, it is also
convenient to invert these equations for the ensuing discussion to obtain
\begin{eqnarray}
T_{\mathrm{res}} &=&4T_{1}T_{2}/(2 l_{\mathrm{cav}}/l_{\mathrm{abs}}\sqrt{\bar{%
\varepsilon}}+T_{1}+T_{2})^{2}, \label{Tres}\\
R_{\mathrm{res}} &=&1-\left( T_{2}+2l_{\mathrm{cav}}/l_{\mathrm{abs}}\sqrt{\bar{\varepsilon}%
}\right) T_{\mathrm{res}}/T_{2}, \label{Rres}\\
T_{1,2} &\equiv&\exp \left[ -\left( l_{\mathrm{cav}}/2\pm \Delta
d\right) /l_{\mathrm{loc}}  \right] ,\label{T12}
\end{eqnarray}
where $\Delta d$ is the shift of the cavity's center from that of the
sample, replacing $l_1$.

Using data on the transmission and reflection coefficients and the
peak widths for the sample shown in Figure \ref{fig_trans} we have
calculated internal parameters of our disordered system.
These results are shown in Table~I.
Shown in  columns 2,3 and 4 are $1-R$, $T$ and $\delta f$, respectively, 
measured for the
resonances indicated in column 1.  We use an extra vertical bar
between columns 4 and 5 to emphasize the fact that the measured resonance
values lie to the left of this double-bar and the retrieved system parameters lie
to the right.
The value of $\Gamma/\sqrt{\epsilon}$
(column 5) has been retrieved via equation \ref{eqn:absorptionlength}. 
The loss tangent $\tan(\alpha) = 
\Gamma c/(2 \pi f \sqrt{{\bar{\epsilon}}}$) is presented in column 6.  The value of
the loss tangent averaged over the five resonances equals $8.35 \times  10^{-4}$.  
The weighted
loss tangent  for our disordered quartz/Teflon system
is $5.2 \times 10^{-4}$, so that the measured and
retrieved values of the absorption agree to within the
accuracy of the experiment.
Thus from external measurements we have retrieved two internal
parameters of the system, the localization length, which relates to
disorder, and the absorption length.

\begin{table}
\begin{tabular}{|c|c|c|c||c|c|}
    \hline
frequency (GHz) &  1-R &   T  & $\delta f$ (GHz) & 
$\Gamma /\sqrt{{\bar \epsilon}} \cdot 10^2$  & $\tan \alpha \cdot 10^4$\\
    \hline
$f_1 = 83.5$   & .978 & .75  & .40  &  .83  & 4.77\\
$f_2 = 92.0$   & .998 & .33  & .39  &  2.6   & 13.45\\
$f_3 = 105.7$   & .993 & .31 & .34   &  2.25 & 10.14\\
$f_4 = 101.8$   & .87 & .18 & .25  &  1.33   & 6.22 \\
$f_5 = 99.8$    & .77   & .30  & .45  &  1.5 & 7.16 \\
    \hline
\end{tabular}
\label{table}
\caption{{\textit Measured and retrieved parameters associated with 5 resonances
shown in Figure \ref{fig_trans}}.  The subscripts on the 
frequencies refer to the peaks indicated in the plot.
The localization length is
1 cm, as obtained from the non-resonant transmission coefficient.}
\end{table}

Finally, we note that numerical simulations were also 
performed with the standard 
propagator matrix method,
as mentioned briefly in  Sec.~\ref{sec:data}, in order to obtain 
information about the electromagnetic
field inside the sample.  For these simulations we used the 
precise experimental 
ordering of quartz and Teflon wafers.

\section{Discussion}
\label{sec:discussion}

\begin{figure}
\leftline{
\includegraphics[width=75mm]{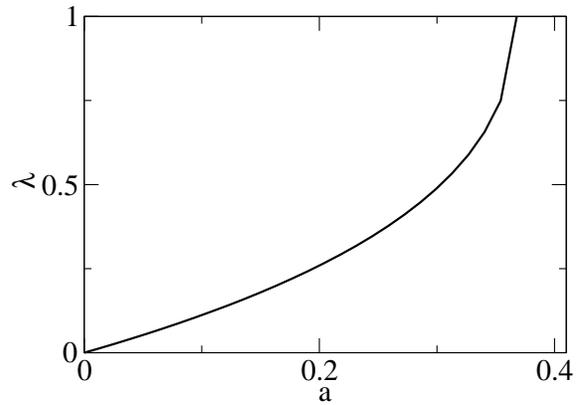}
}
\caption{Shown is the solution to the transcendental Eq.~(\ref{eqn:rescale}), 
where $\lambda = l_{\mathrm{cav}}/2l_{\mathrm{loc}}$ and
$a$ is given in Eq.~(\ref{eqn:adefn}).}
\label{fig:cavitylength}
\end{figure}

Equation~(\ref{eqn:cavitylength}) requires further analysis, as it is transcendental.  
We can rescale it as
\begin{equation}
\lambda=a \exp(\lambda)\,,\label{eqn:rescale}
\end{equation}
where \begin{eqnarray}
\lambda &\equiv& l_{\mathrm{cav}}/2l_{\mathrm{loc}}\,,\\
a&\equiv& \frac{c}{2\,l_{\mathrm{loc}}\,\bar{\varepsilon}\,\delta\omega\,\sqrt{T_{\mathrm{res}}}}
\exp(-L/2l_{\mathrm{loc}})\,.\label{eqn:adefn}
\end{eqnarray}
Equation~(\ref{eqn:rescale}) has a real solution in terms of a 
ProductLog for $a\in(-\infty,1/e]$:
\begin{equation}\lambda=- \mathrm{ProductLog}(-a)\,,\end{equation} a
special function which can be obtained numerically. 
 
In the physical
region of $a\in[0,1/e]$, $\lambda\rightarrow 1$ as $a \rightarrow
1/e$.  Since the localization length must be smaller than the
cavity size, it follows that $\lambda$ is confined to the relatively 
narrow range of $\frac{1}{2}e^{-\frac{1}{2}} \approx 0.3$ and $e^{-1}
\approx 0.37$. 
Then the resonance width $\delta\omega$ must fall off
exponentially as the system size is increased.  
%But this is just what
%we expect for localized states: they are rare, and their density
%amidst all possible states falls to zero as the system size grows to
%infinity.

\begin{figure}
\begin{center}
    \includegraphics[width=90mm]{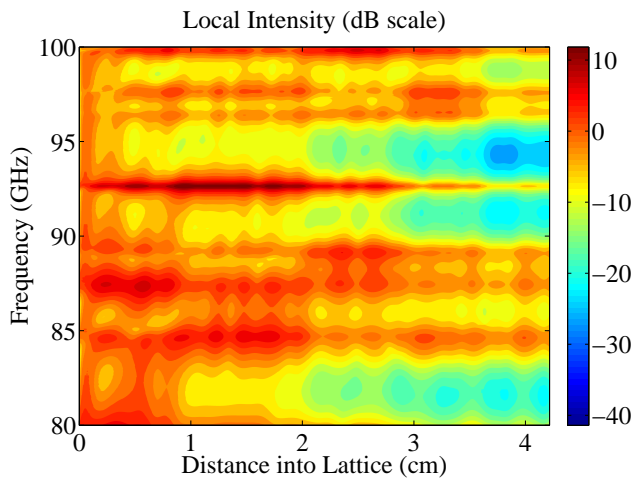}
    \caption{\label{fig_localization}
    \textit{Localized  E field within the sample.}
    Shown is a plot of the calculated E-field intensity (in
    $dB$) within the
    scattering medium. By comparing the frequencies with those in Fig.~\ref%
    {fig_trans} one can see the strong localization of the field at
    transmission resonances. }
\end{center}
\end{figure}

\begin{figure}
\centerline{
    \includegraphics[width=100mm]{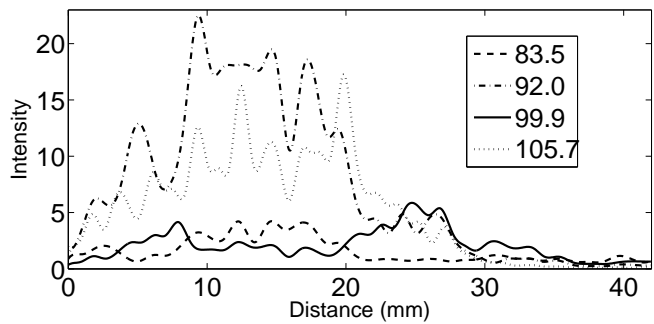}}
    \caption{\label{fig_3peaks}
    \textit{Localized modes at regions of enhanced
    absorption.}  E-field intensity
    as a function of space inside the sample at four resonant frequencies
    used in the analysis in Table I.  Localization
    results in a varying degrees of field enhancement inside the random
    effective cavity.  The stronger the field enhancement, the greater the
  path-length-induced absorption.}
\end{figure}

Figure~\ref{fig_localization} shows the calculated spatial
distribution of the E-field inside the medium for the same sample as
was used in Fig.~\ref{fig_trans}, for both resonant and non-resonant
frequencies.  
Figure \ref{fig_3peaks} shows the spatial distribution of energy at three 
resonances.

At resonance, the energy density within the localized mode, i.e., inside
the effective cavity,
can be orders
of magnitude larger than that of the incident wave.  This huge
field enhancement has many potential applications.  For example, at
optical frequencies, this increased energy density has been
exploited to produce a random laser~\cite{lasers} based on a layered
(1D) medium. 
Off resonance,
the dielectric multi-layer
presents an almost perfect reflector; it is only the exponential
tail of the field that penetrates the sample.
%\emph{\ (We can add
%more possible applications here)}

\begin{figure*}
\centerline{
\includegraphics[width=130mm]{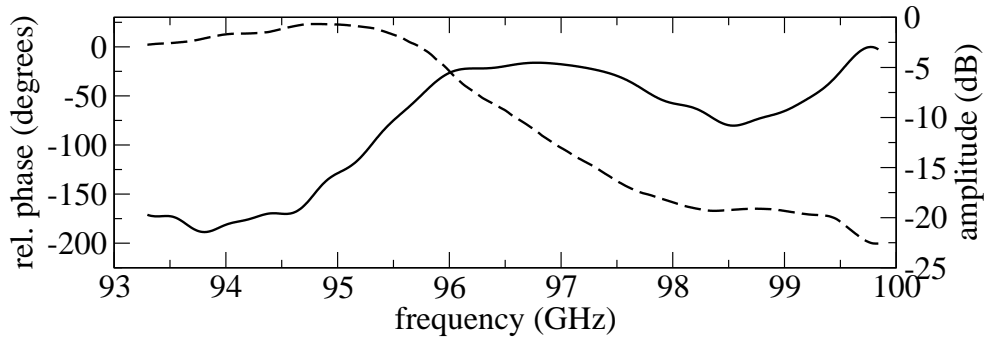}
}
\caption{In order to interpret the fast and slow-light features of our
measurement, we zoom in on a typical transmission resonance.
For this purpose we could choose any of the resonances
but here we zoom in around 97 GHz.
We show the phase (dashed curve) and amplitude (solid
curve) of the E-field transmitted through the same sample as in
Figs.~\ref{fig_trans} and~\ref{fig_localization}, as a function of
frequency  $f$ for the transmission resonance at 97 GHz.
In band gaps the phase is nearly stationary while
at resonance the inverse phase derivative,  $(d\phi(f) /df)^{-1}$
decreases markedly.}
\label{fig_amp_phase}
\end{figure*}

Slow light is associated with strong dispersion and can be seen in
situations involving resonant transmission, for example, with defect
states in a band gap~\cite{Bigelow07112003} or with ultracold atomic
gases~\cite{Hau}.  In the case of localization, slow light and
enhanced absorption go hand in hand because they have the same
physical origin -- both of them are caused by multiple scattering,
which increases dramatically the photon path and the resonance dwell
time.  Plotted in Fig.~\ref{fig_amp_phase} is the measured frequency
dependence of the phase, $\phi (f)$, of the signal transmitted
through a random sample.  At each resonance there is a pronounced
decrease of the inverse phase derivative, $(d\phi (f
)/df)^{-1}$.  If there were a well defined group, i.e., a
pulsed measurement, and if there were a well defined
local wave number, then the inverse phase derivative would be
proportional to the group velocity.  For example, one could divide
the transmitted phase by the thickness of the sample to get an
average wave number.  Such an analysis, however rough, gives group
velocities on the order of 10\% c.  But our measurement doesn't have
sufficient bandwidth to synthesize pulses small compared to the size
of the sample. 

The corresponding time delay associated with a resonance can be
estimated to be 
\begin{equation}
\Delta t\propto \frac{1}{\Delta
f}=\frac{l_{\mathrm{cav}}}{c\,T_{\mathrm{typ}}} \gg \frac{l_{\mathrm{cav}}}{c}.
\end{equation}
%where $\Delta \nu $ is the line-width of the  resonance.
%, $l_{\mathrm{cav}}$ is the width of the effective resonance cavity,
%and $T_{\mathrm{typ}}$ is the transmission coefficient for a typical
%frequency. 
Taking $l_{\mathrm{cav}}$ of the order of twice the localization
length,
$2\times 1$cm,
and $T_{\mathrm{typ}}\simeq 0.01$, we obtain $\Delta t\propto
l_{\mathrm{cav}}/c\times 1/T_{\mathrm{typ}}=2\:\mathrm{cm}\times
100/(3\times 10^{10}\mathrm{cm}/\mathrm{s}) \approx 7$ ns. If we were
speaking of well-defined pulses, such a delay would correspond to a
group velocity an order of magnitude less than
the speed of light. Although our
measurements are fundamentally frequency-domain, the
structure of the spectral peaks of the transmission coefficient can
be related to the conventional notion of slow light in this sense.

\section{Conclusions}
\label{sec:conclusions}

We have developed an experimental system for studying
disorder-induced wave phenomena in dielectrics at millimeter wave
frequencies. In our system we see localization even in samples that
are only  four times the localization length. The
deep resonances connected with localization allow one to observe
enhanced absorption and 
disorder-induced delays that can be associated with group
velocities much less than the speed of light in vacuum. Furthermore,
off-resonance, the random dielectric stack
becomes a near perfect reflector. This gives rise to a transmitted
E-field phase that is nearly independent of frequency at the band-edge.
We have interpreted the experimental data in terms of an effective cavity
model, which enables us to retrieve the localization and absorption lengths
from frequency dependent reflection and transmission coefficients.

This material is based upon work supported by the National Science
Foundation under Grants EAR-0337379, EAR-041292 and PHY-0547845.

\end{document}